# Design parameter space for a
# High Pressure Optimized Dense Plasma Focus operating with Deuterium

S K H Auluck
HiQ TechKnowWorks Pvt. Ltd,
Navi Mumbai, India

## Abstract

The potential of the Dense Plasma Focus (DPF) for industrial applications in many fields is well recognized, although yet to be realized in practice. Particularly attractive is the possibility of its use as inexpensive industrial source of nuclear reactions for diverse high value applications such as fast pulsed neutron radiography of hydrogenous materials, non-intrusive neutron interrogation of concealed organic contraband and rapid production of short lived radioisotopes for medical diagnostics and therapy. Recently, it has been suggested that it may be possible to operate the DPF efficiently in a High-Pressure-Optimized (HPO) mode. This paper explores the design parameter space for such HPO-DPF based on the revised Resistive Gratton-Vargas (RGV) model with a view to identify a practicable set of system parameters and their scaling. The current waveform predicted by the revised RGV model for the chosen set of parameters is fitted to the Lee model to estimate the likely neutron yield.

I. Introduction:

The potential of the Dense Plasma Focus (DPF) for commercial applications in diverse fields is well recognized [1,2,3], although yet to be realized in practice. Particularly attractive is the possibility of its use as inexpensive industrial source of nuclear reactions for high value applications such as fast pulsed neutron radiography[4] of hydrogenous materials (like rocket motors), non-intrusive neutron interrogation of concealed organic contraband and explosives [5] and rapid production of short lived radioisotopes for medical diagnostics[6,7] and therapy [1]. These applications seek to utilize the fact that DPF has an unusually high binary nuclear reaction rate caused by presence of an intense stream of high energy ions trapped[8,9] within the plasma by virtue of its peculiar magnetic field configuration. While no single mechanism has been conclusively established as a definitive cause of this phenomenon, available evidence points towards a process of reconfiguration of the magnetic field [9].

Recently, it has been suggested [10] that it may be possible to operate the DPF *efficiently* in a High-Pressure-Optimized (HPO) mode, with deuterium gas pressure one or two orders greater than the current practice of operating at few (<10) mbar, assuring that a pre-specified fraction of stored energy gets transformed into magnetic energy and no energy remains in the capacitor bank at the time of the current derivative singularity. Judging by the experience of the two DPF devices (NX2 at Singapore [11] and FF1 at Lawrenceville[12], USA) which have reported significantly higher neutron yield than devices of comparable energy by operating at higher than 10 mbar deuterium pressure, and numerical simulations with the Lee model at 60 torr [13] which also show increased neutron yield, one could expect orders of magnitude improvement in the yield of fusion neutrons (or other binary reactions) if *efficient* DPF operation could be achieved at deuterium pressures of 0.1-1 bar. This paper explores the design parameter space for such HPO-DPF operating with deuterium using the revised Resistive Gratton-Vargas (RGV) model [10,14,15] with a view to identify a practicable set of system parameters and their scaling.

The Resistive Gratton-Vargas Model belongs to the class of oversimplified models of the DPF, which are based on assumptions well-known to be not valid in practice but yet have the ability to fit experimental current waveforms [16] by adjusting some model parameters. The famous Lee model [13,17,18,19] is another member of this class. While the Lee model solves its differential equations numerically, the RGV model does it analytically, producing algebraic relations rather than tables of numbers. The RGV model *has no in-built physics* other than an assumption of equality between the magnetic pressure of azimuthal magnetic field driving the plasma and the "wind pressure" resisting its motion at an imaginary surface that plays a role similar to the imaginary center of mass in mechanics. It produces an analytical formula for the current waveform and gives relations about the partition of stored energy among magnetic energy, capacitive energy, work done and resistive loss. The Lee model contains, besides the snowplow hypothesis, shock dynamics, slug model, pinch effect, thermodynamics including radiation loss and anomalous resistivity and produces an estimate of most kinds of emissions from the plasma made with many different kinds of gases, provided some model fitting parameters are chosen in such manner as to fit a given current waveform. It also takes into account a device scaling parameter called the

Speed Factor $S \equiv I/(a\sqrt{\rho_0})$, which is found [20] to have the value 89±8 kA/cm per (torr)$^{1/2}$ for many existing neutron-optimized plasma focus devices.

This paper *utilizes both models* to investigate the possibility of designing DPF *working efficiently* at a gas fill pressure in the range of 0.1 bar to 1 bar of deuterium, as a first step towards realization of a small plasma device that can generate nuclear reaction products in quantities of commercial interest.

Traditionally, DPF has been approached from the point of view of an object of scientific study, where the interest is in relating its observed phenomenology to known physics using a variety of experimental and theoretical techniques. In this approach the device is a 'given': questions about choice of its operating parameters are never part of the discussion. This is because the operating parameters are chosen based on practical issues such as the level of funding, availability of off-the-shelf capacitors, cables and spark gaps and access to relevant technical expertise. In this approach, the construction and operation of the device is based on empirical thumb rules. These rules-of-thumb represent accumulated knowledge from efforts by many groups over several decades for systematic variation of parameters. However, these efforts have never had the benefits of a scaling theory comparable in scope to the Lee model or the RGV model. As a result, some operating parameters, such as plasma density, tend to be nearly identical [21] over a wide range of stored energy. Based on available literature, it is not clear whether this near constancy of density is a phenomenon rooted in fundamental physics or an artifact of empirical research processes.

This paper looks at the DPF from a diametrically opposite point of view, where the interest is in commercial utilization of DPF as an economical producer of nuclear reactions. The device is therefore not a 'given': the choice of its design must necessarily be guided by maximization of its economic potential. Rather than make any device and see what information can be extracted from it, this approach would seek to develop a design logic that utilizes existing knowledgebase to connect user requirements with initial prototype system specifications, which can be refined empirically, and *then* study the behavior of the device. The economic stakes are assumed to be large enough to ensure that availability of funding, relevant expertise and off-the-shelf components are not insurmountable obstacles.

This paper attempts construction of such design logic and its illustration with concrete examples using the following components of available knowledgebase:

1. Plasma focus dynamics is qualitatively independent of the scale of the device [21].

2. The RGV model is capable of reproducing experimental current waveforms using static inductance, static resistance and pressure as fit parameters [16].

3. The Lee model is capable of providing estimates of nuclear reaction rates [18] using an empirically-calibrated phenomenological beam target model (which ignores experimental evidence [22] concerning occurrence of toroidally streaming energetic ions) provided the fit parameters are chosen to reproduce a given current waveform.

4. Conservation laws provide a lower limit on the velocity of the plasma current sheath [23].

5. Local electric field necessary for electric breakdown in neutral hydrogen gas can be estimated from the electron impact ionization cross-section and total electron collision cross-section of molecular hydrogen and is proportional to the density of gas.[23]

Specifically excluded from this discussion are empirical thumb rules. This is justified on the grounds that past efforts have not had the benefits of scaling theories comparable in scope to the RGV model or the Lee model. Commonly ignored experimental evidence [22] from many reputed laboratories concerning existence of axial magnetic field and toroidally streaming energetic ions in the DPF as well as the theoretical demonstration [22] that they are natural consequences of conservation laws also indicate that time is ripe for DPF research to evolve in novel ways which may be of great practical value as well. This paper briefly discusses the relevance of these phenomena to its theme of high pressure optimized operation.

This is not merely an engineering exercise but a vital aspect of fundamental studies in fusion plasma physics, that has been neglected for far too long. Fusion reactor physics came to symbolize tokamak physics only because scaling theories of tokamak experiments enabled construction of a hierarchy of machines of increasing complexity, in spite of the fact that theoretical models failed to explain crucial aspects of tokamak physics [24]. In contrast, plasma focus (and most other early fusion reactor concepts) did not have a scaling theory which could enable design and construction of a device for achievement of a desired level of current at the pinch time. This paper attempts to provide a scaling theory, enabling future construction of machines operating in high pressure optimized mode. The scheme of constructing dimensionless conservation law equations [22] for the DPF predicts that the plasma density in the pinch phase should increase when the plasma focus operates *efficiently* at high fill density of deuterium. It also predicts occurrence of toroidally streaming energetic deuterium ions[22] without the mediation of instabilities or anomalous resistivity. If these two predictions are experimentally borne out in the case of HPO-DPF, the nuclear reaction rate may increase by orders of magnitude over that predicted by traditional models [18].

The next section II recapitulates some aspects of the RGV model and physics external to the RGV model relevant to the present discussion. Section III constructs a design logic, proposes an optimization algorithm that explicitly includes a high pressure of deuterium (in the 0.1-1 bar range), defines the operating parameter space in functional terms and looks at the system scaling behavior. Section IV illustrates the proposed algorithm using two concrete examples, obtains the current waveforms predicted for parameter sets for these examples from RGV model, fits the Lee model to the current waveforms and looks at the predictions of the neutron yield from the Lee model. Section V discusses some aspects of scaling of neutron yield estimates. Section VI summarizes and concludes the paper.

II. <u>Some results from the RGV model and from physics external to it</u>

This section recapitulates some results of the RGV model necessary for this discussion without repeating their basis or nomenclature, which may be found elsewhere[10,14,15,16]. Also mentioned are some results from physics external to the RGV model [23].

The RGV model maps the ten parameters describing a dense plasma focus (voltage $V_0$, capacitance $C_0$, static inductance $L_0$, static resistance $R_0$, mass density $\rho_0$, anode radius a, insulator radius $\mathbb{R}_I$, cathode radius $\mathbb{R}_C$, anode length $\mathbb{L}_A$ and insulator length $\mathbb{L}_I$) on to 7 independent dimensionless model parameters, which completely determine the fractions of energy converted into magnetic energy and work done, remaining in the capacitor bank and that resistively dissipated. Time t is represented in dimensionless form by a variable $\tau$ proportional to the charge that has flown in time t:

$$\tau(t) = \frac{1}{Q_m} \int_0^t I(t') dt' \quad ; \quad Q_m \equiv \mu_0^{-1} \pi a^2 \sqrt{2\mu_0 \rho_0} \qquad 1$$

The anode radius serves as the length scale and four of the seven dimensionless parameters represent the scaled geometry of the device: $\tilde{z}_A \equiv \mathbb{L}_A/a$, $\tilde{z}_I \equiv \mathbb{L}_I/a$, $\tilde{r}_I \equiv \mathbb{R}_I/a$, $\tilde{r}_c \equiv \mathbb{R}_C/a$. Of the remaining three, $\gamma \equiv R_0 \sqrt{C_0/L_0}$ governs the resistive dissipation. The two parameters $\varepsilon \equiv Q_m/C_0 V_0 = (\mu_0 C_0 V_0)^{-1} \pi a^2 \sqrt{2\mu_0 \rho_0}$, and $\kappa \equiv \mu_0 a / 2\pi L_0$ involve the mass density of the fill gas $\rho_0$ and the physical scale of the device 'a'.

The analytic structure of the RGV model allows [10] mathematical formulation of the following two operational definitions of an optimized DPF facility in terms of the dimensionless model parameters:

1. The fraction of energy remaining in the capacitor bank at the instant of the current derivative singularity (CDS), should be zero.

$$\varepsilon^{-1} = \tau_p \equiv 2(\tilde{z}_A - \tilde{z}_I) + 1 \qquad 2$$

2. The fraction of energy converted into magnetic energy at the time of CDS should equal a specified "nominal design efficiency $\eta_0$" (~ 0.7-0.75). This translates to

$$\kappa^{-1} = \frac{\mathcal{L}(\tau_p)(\eta_0 - 2A)}{(1 - \eta_0)} ; A \equiv \left( \frac{m_0(\tau_p)}{\mathcal{L}(\tau_p) \tau_p} - \frac{m_1(\tau_p)}{\mathcal{L}(\tau_p) \tau_p^2} \right) \qquad 3$$

where

$$\mathcal{L}(\tau) = 0.5\tilde{z}_I \text{Log}(\tilde{r}_I^2 + \tau) + k_1 \text{Log}(\tilde{r}_C)\tau^{1.5} \quad 0 < \tau \leq \tau_{LIFTOFF}$$

$$= \mathcal{L}(\tau_{LIFTOFF}) + \frac{1}{2}(\tau - \tau_{LIFTOFF})\text{Log}(\tilde{r}_C) + k_2\text{Log}(\tilde{r}_C) \quad \tau_{LIFTOFF} \leq \tau \leq \tau_r \quad 4$$

$$= \mathcal{L}(\tau_r) - k_3\text{Log}(\tau_R + 1.00439 - \tau) + 0.00439 k_3 \quad \tau_r < \tau \leq \tau_p$$

$$k_1 = \frac{\lambda_0}{\tilde{r}_C + \lambda_1}; k_2 = \lambda_2 + \lambda_3\tilde{r}_C + \lambda_4\tilde{r}_C^2; k_3 = \lambda_5 + \lambda_6\tilde{r}_C + \lambda_7\tilde{r}_C^2; \quad \tau_{LIFTOFF} = \tilde{r}_C^2 - \tilde{r}_I^2; \quad \tau_r = 2(\tilde{z}_A - \tilde{z}_I);$$

$\tau_p = \tau_r + 1$; $\lambda_0 = 0.276304$; $\lambda_1 = -0.68924$; $\lambda_2 = -0.08367$; $\lambda_3 = 0.105717$; $\lambda_4 = -0.02786$;

$\lambda_5 = -0.05657$; $\lambda_6 = 0.263374$; $\lambda_7 = -0.04005$.

$$m_0(\tau) \equiv \int_0^\tau d\tau' \mathcal{L}(\tau'); m_1(\tau) \equiv \int_0^\tau d\tau' \tau' \mathcal{L}(\tau') \quad 5$$

The instant of CDS, also popularly known as "the pinch time", is denoted by $\tau_p$

This algorithmic definition of DPF optimization is by itself not sufficient to determine all the 10 parameters of a DPF facility; nevertheless, it limits the design parameter space in such manner that if a facility satisfies these criteria, it is sure to transfer maximum energy to the plasma at the instant of the CDS according to the RGV model.

An additional relation has been derived [10] from the demonstration of a lower bound $v_L$ (~7.9x10$^4$ m/s for deuterium gas) on the plasma current sheath velocity [23]:

$$v_{sp0} \equiv \frac{\mu_0 I_0}{2\pi a\sqrt{2\mu_0 \rho_0}} = fv_L, \quad f > 1, \quad I_0 \equiv \frac{V_0}{Z_0}, \quad Z_0 \equiv \sqrt{\frac{L_0}{C_0}} \quad 6$$

It has also been demonstrated [23] that the local electric field necessary to cause electric breakdown in neutral hydrogen gas can be estimated from the electron impact ionization cross-section and total electron collision cross-section of neutral molecular hydrogen and is given by $E_b = \upsilon n_0, \upsilon \simeq 8.21 \times 10^{-18} \cdot \text{Volt-m}^2$ for hydrogen.

III. Construction of design logic:

This section presents a logical process for determining all 10 parameters of an initial prototype of DPF facility starting from end-user requirements, which include a deuterium gas fill pressure in the range of 0.1-1 bar, and illustrates it using concrete examples. This set of parameters can then be refined using the RGV model, further examined using the Lee model and then refined empirically by actually manufacturing the device.

This process begins by recognizing that the resistance of the circuit can be neglected in the initial estimate, putting $\gamma \equiv R_0 \sqrt{C_0/L_0} \approx 0$. Next, one recognizes that insulator thickness is usually negligible compared with its internal radius, which is quite close to the anode radius. This allows the approximation $\tilde{r}_I \equiv \mathbb{R}_I/a \approx 1$. Similarly, one can choose $\tilde{r}_C \equiv \mathbb{R}_C/a \approx 1.5$, noting that the formula for inductance is valid only for the range $\tilde{r}_I + 0.2 \leq \tilde{r}_C \leq 2.0$ [14] and this value would anyway be revised in the next iteration. Out of the 10 parameters of the facility comprising 7 parameters of the RGV model and 3 degrees of freedom, 3 are thus fixed by this initial choice. Equations 2, 3 and 6 provide 3 independent relations. The operating pressure is required to be in the range 0.1-1 bar, so that must be retained as a discretionary parameter. Voltage $V_0$ and inductance $L_0$ must be examined by the capacitor bank designer for feasibility and therefore must also be discretionary parameters. Thus 9 resources (3 initial estimates, 3 relations and 3 discretionary parameters) are available; in order to determine 10 parameters one additional relation is therefore necessary. This paper provides that relation.

The additional relation can be derived from the condition that at the instant of capacitor bank switching, the electric field across the insulator should be greater than the estimated breakdown strength mentioned at the end of Section II.

$$E_{ins} = \frac{V_0}{a\tilde{z}_I} = gE_b, g > 1 \qquad 7$$

The algorithm for determining initial prototype specifications can thus be summarized in terms of the following *algebraic relations*

$$\tilde{r}_I = 1., \quad \tilde{r}_c = 1.5, \quad \gamma = 0, \qquad \text{initial choices} \qquad 8$$

$$2(\tilde{z}_A - \tilde{z}_I) + 1 = \frac{\mu_0 C_0 V_0}{\pi a^2 \sqrt{2\mu_0 \rho_0}} \qquad \text{zero stored energy at CDS} \qquad 9$$

$$\frac{\mathcal{L}(\tau_p)(\eta_0 - 2A)}{(1-\eta_0)} = \frac{2\pi L_0}{\mu_0 a} \qquad \text{fraction } \eta_0 \text{ of energy converted to magnetic energy} \qquad 10$$

$$\frac{V_0}{a\tilde{z}_I} = g\chi\rho_0, g > 1$$

$$\chi \equiv \frac{\upsilon}{m_{deu}} = 2.46 \times 10^9 \cdot \text{Volt-m}^2/\text{kg for deuterium} \qquad \text{electric breakdown at insulator} \qquad 11$$

$$\frac{\mu_0 V_0}{2\pi a \sqrt{2\mu_0 \rho_0}} \sqrt{\frac{C_0}{L_0}} = fv_L, \quad f > 1 \qquad \text{lower bound on sheath velocity} \qquad 12$$

$$\rho_0 = 0.17846 \left(\text{kg/m}^3\right) p\left(\text{bar D}_2\right) \qquad 13$$

*where p is to be chosen in the range 0.1-1 bar as a user requirement.*

$V_0$ and $L_0$ are to be decided on the basis of overall feasibility. This involves the decision about going for a bank of parallel capacitor-spark-gap modules or parallel Marx banks, adopting cable connections or strip-line connections, and decisions about electro-mechanical construction of the interface between the DPF electrodes and the power source, which must not only provide a low inductance, high voltage, high current interface with the capacitive energy storage but also needs to withstand high magnetic stress, resist electrode erosion because of high surface current density and provide an adequate vacuum seal. The role of the RGV model lies in providing first order estimates of anode radius a, cathode radius $\mathbb{R}_C$, anode length $\mathbb{L}_A$, insulator length $\mathbb{L}_I$, circuit current at the time of current derivative singularity (referred as the pinch current) $I_p$ and stored energy $E_0$ as a function of the 3 discretionary parameters voltage $V_0$, static inductance $L_0$, and pressure p. The following survey of parameter space, with $V_0$ in the range 10-200 kV, $L_0$ in the range 10-50 nH, p 0.1-1 bar, nominal design efficiency $\eta_0 \approx 0.7$, f=1.2, g=1.5, illustrates the nature of technical decisions.

Fig 1 shows that the scaled "pinch" current $\tilde{I}_p$ (current at CDS normalized to $I_0 \equiv V_0/Z_0$) is almost constant: (varying between 0.62 to 0.68)

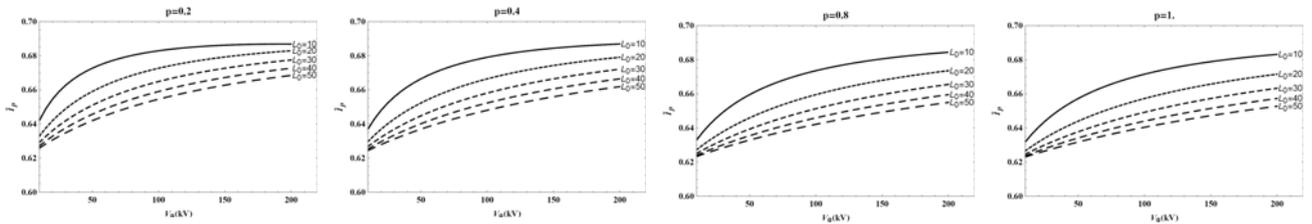

Fig 1: Variation of scaled pinch current $\tilde{I}_p$ as a function of $V_0$, $L_0$ and p

The absolute pinch current and stored energy are related as

$$I_p = \tilde{I}_p \frac{V_0}{Z_0}; E_0 = \tfrac{1}{2} \frac{V_0^2 L_0}{Z_0^2}; I_p^2 = \frac{\tilde{I}_p^2 E_0}{\tfrac{1}{2} L_0} \qquad 14$$

The observed constancy of $\tilde{I}_p$ in the optimization algorithm therefore implies that the square of the pinch current scales as the stored energy; *there is no saturation effect for current with energy as observed by Lee [18,19]*. This is because the algorithm prescribes an optimum value of the impedance $Z_0$ for every choice of the three discretionary parameters, unlike the argument presented by Lee[18,19]. So the first take-away of this exercise is that the observed failure of neutron yield above a certain energy threshold *may not be* because of failure of energy transfer from the capacitor bank to the plasma as suggested by Lee[18,19]; the proposed algorithm guarantees efficient energy transfer to the plasma. The question of failure of neutron yield at high currents therefore continues to remain open for the moment.

Fig 2 shows that the optimum value of impedance $Z_0$ is linearly dependent on $V_0$, with parameters which are themselves linearly dependent on $L_0$ with parameters which are in turn functions of p.

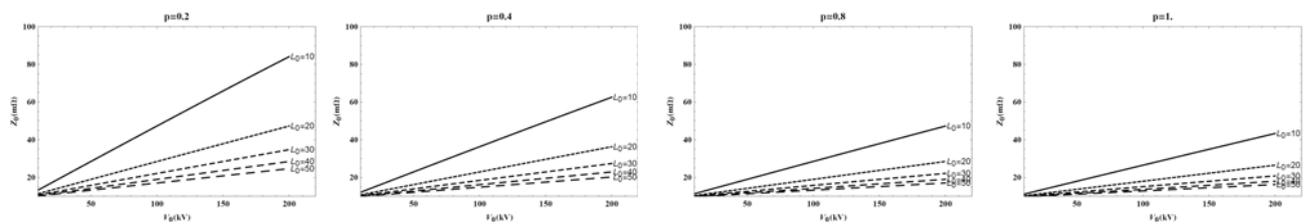

Fig 2: Variation of impedance $Z_0$ as a function of $V_0$, $L_0$ and p

This implies that when the designer chooses "practicable" values of voltage and inductance (say 20 kV and 20 nH) for a proof-of-concept[26] of high pressure operation of DPF at a certain pressure (say 0.2 bar), the stored energy (which forms the major component of the cost) is already decided (22.86 kJ). Figure 2 shows that opting for a higher pressure would involve lower impedance and hence a higher energy (and cost).

On the other hand, an industrial scale DPF should try to maximize the "bang-for-the-buck (BB)". Going with the traditional $I_p^4$ scaling of fusion reaction yield and assuming the cost to scale proportional to stored energy, this requires maximization of the "cost effectiveness parameter"

$$BB \equiv \frac{I_p^4}{E_0} = 2\frac{\tilde{I}_p^4 V_0^2}{Z_0^2 L_0} \qquad\qquad 15$$

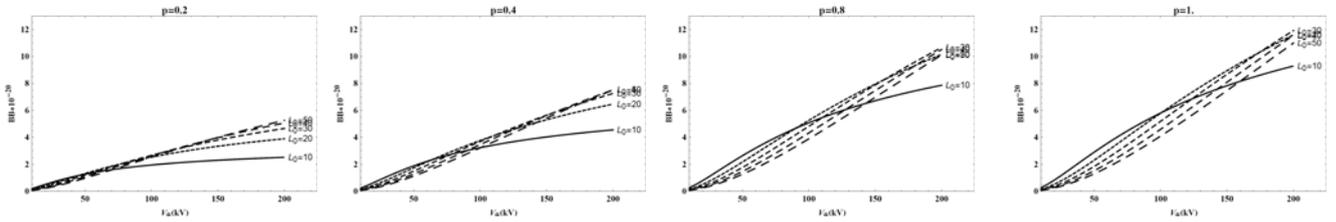

Fig 3: Cost effectiveness parameter BB as a function of $V_0$, $L_0$ and p

Surprisingly, this indicates that higher pressure is more cost effective. This result follows from the above described algorithm, which does not include any possible benefits of higher binary nuclear reaction rate because of increased density. Fig. 3 also suggests that excessive efforts at lowering the static inductance may not be warranted: inductance of 40 nH is preferable to 10 nH. The benefits of choosing a higher working voltage are also evident. An industrial facility therefore needs to follow a construction strategy similar to SPEED-II [25].

The following figures present the results for anode radius, anode length, insulator length.

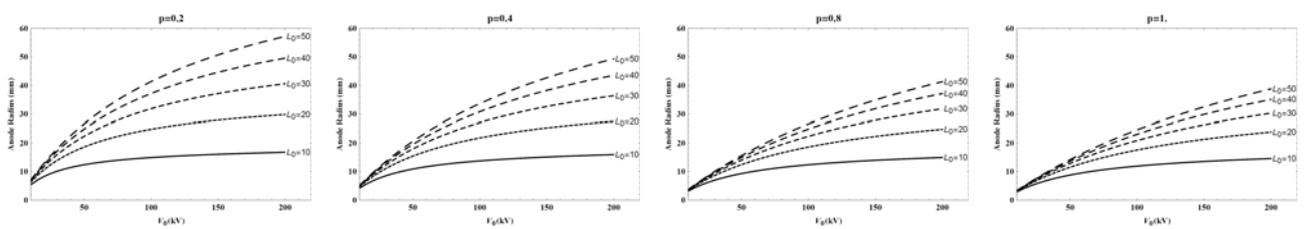

Fig 4: Anode radius in mm as a function of $V_0$, $L_0$ and p.

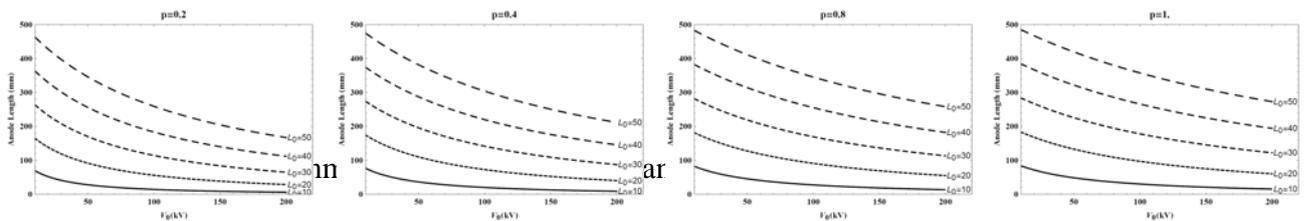

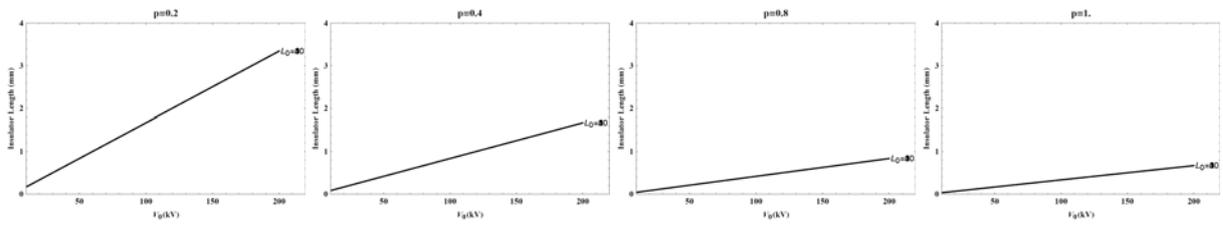

Fig 6: Insulator length in mm as a function of $V_0$, $L_0$ and p

The surprise is that the optimized insulator length revealed by the algorithm is less than 4 mm, *a result completely contrary to research experience till now* (providing post-facto justification for ignoring empirical thumb rules). The insulator length optimized for deuterium is found to be independent of inductance and well represented by the formula:

$$\mathbb{L}_I(\text{mm}) = 0.00334164 \frac{V_0(\text{kV})}{p(\text{bar})} \qquad 16$$

For the sake of completeness, the following figures show the stored energy, pinch current and capacitance as a function of the discretionary variables.

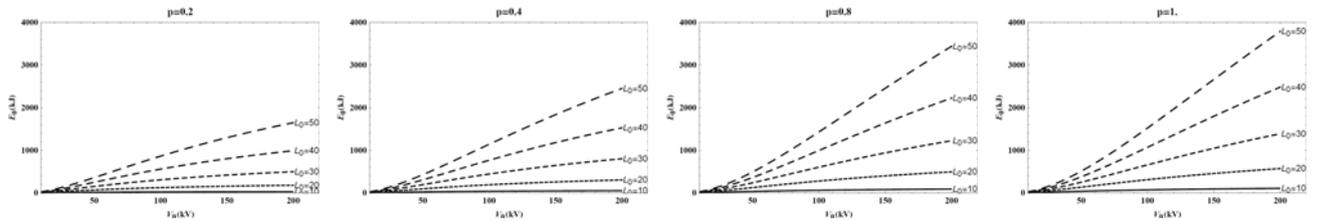

Fig 7: Stored energy in kJ as a function of $V_0$, $L_0$ and p

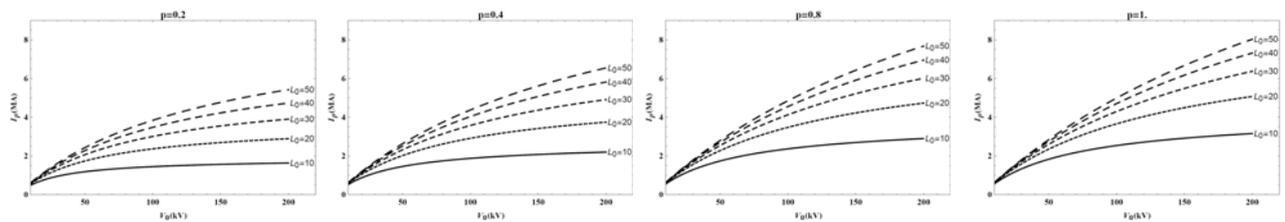

Fig 8: Pinch current in mega-amperes as a function of $V_0$, $L_0$ and p. Note that this is for a zero resistance case for the first stage of design iteration; it only gives a ball park figure.

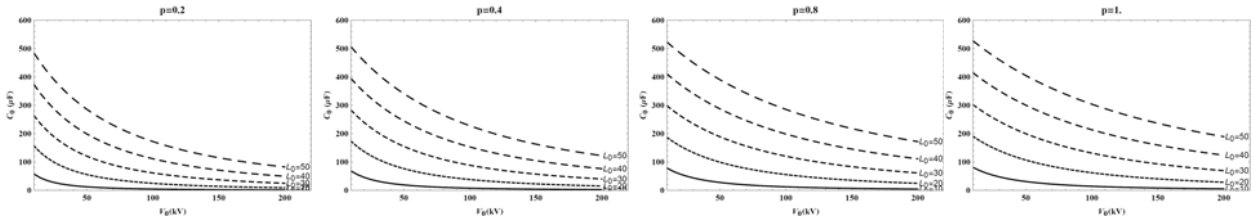

Fig 9: Capacitance in µF as a function of $V_0$, $L_0$ and p

The main outcome of the above exercise is the demonstration that the proposed algorithm is capable of completely specifying the first order prototype design iteration parameters for application-driven choices of the three discretionary parameters $L_0$, $V_0$ and p over a wide range in energy / current, without taking recourse to any empirical thumb rules.

The next section IV illustrates the design iteration procedure using two concrete examples.

IV. Numerical examples:

This section illustrates the use of the design logic and optimization algorithm presented in section III using two examples. The first example, designated PC, corresponds to the case when a proof-of-concept[26] of the idea that high pressure optimized DPF operation is feasible is sought with minimum effort and cost, as a necessary exercise to justify incurring the expense of a full-scope industrial scale facility. The second example, designated IS, corresponds to an industrial scale facility, chosen to be cost effective in terms of parameter BB defined in 15. The discretionary variables for the first case are chosen to be $V_0$=20 kV, $L_0$=20 nH, p=0.1 bar of deuterium. Higher values would lead to higher stored energy and thus higher cost. Lower value of inductance and voltage would complicate construction and switching and require extra efforts and possibly some trial and error again increasing cost. The discretionary variables for the second case are chosen to be $V_0$=200 kV, $L_0$=40 nH, p=1 bar of deuterium, taking cue from fig. 3, which shows that BB is nearly same for 20 nH, 30 nH and 40 nH at 200 kV. Since the voltage indicates that the configuration of the energy storage should be many Marx banks in parallel, opting for a higher inductance would mean lesser number of Marx banks are required to be put in parallel, reducing the number of spark gaps and trigger generators thereby increasing reliability and decreasing cost. Table I lists the parameters obtained from the above algorithm.

Table I: Optimized parameters revealed by proposed algorithm

| Parameters | PC | IS |
|---|---|---|
| Inductance $L_0 (nH)$ | 20 | 40 |
| Voltage $V_0 (kV)$ | 20 | 200 |
| Pressure $p (bar\, D_2)$ | 0.1 | 1 |
| Energy $E_0 (kJ)$ | 18.3 | 2479 |
| Capacitance $C_0 (\mu F)$ | 91.3 | 124 |
| Anode radius $a (mm)$ | 13.5 | 35 |
| Anode length $\mathbb{L}_A (mm)$ | 122.1 | 194.2 |
| Cathode radius $\mathbb{R}_C (mm)$ | 20.2 | 52.6 |
| Insulator length $\mathbb{L}_I (mm)$ | 0.67 | 0.67 |
| Pinch current $I_p (MA)$ | 0.875 | 7.3 |
| Drive parameter $(kA/cm)/\sqrt{torr}$ | 81.6 | 85 |

Fig. 10 and 11 show the results from RGV model calculation[16] for the PC system.

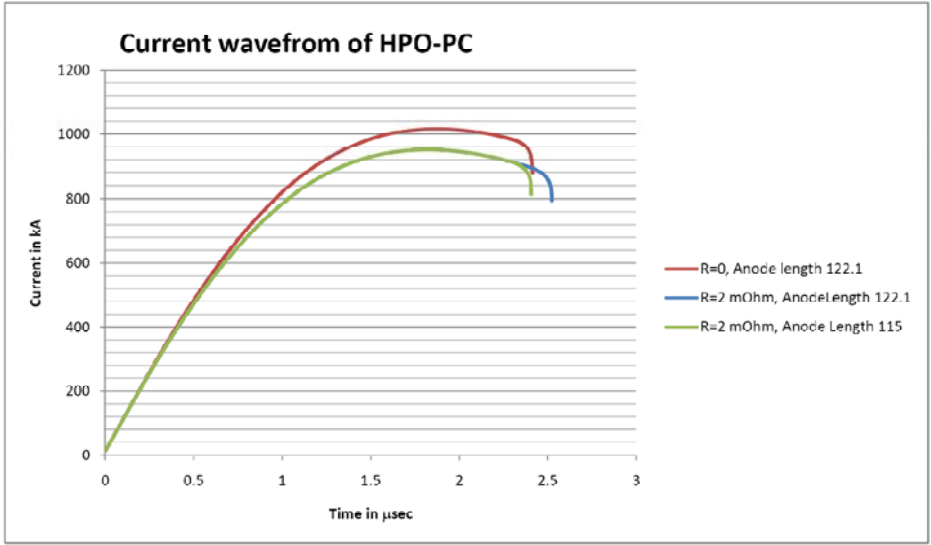

Fig 10: Current waveform calculated from RGV model[16] for proposed High Pressure Optimized Proof-of-Concept system. Initially, resistance is assumed to be zero. Assuming a more realistic resistance value of 2 mΩ leads to a delay in CDS. To compensate, anode length is reduced by trial.

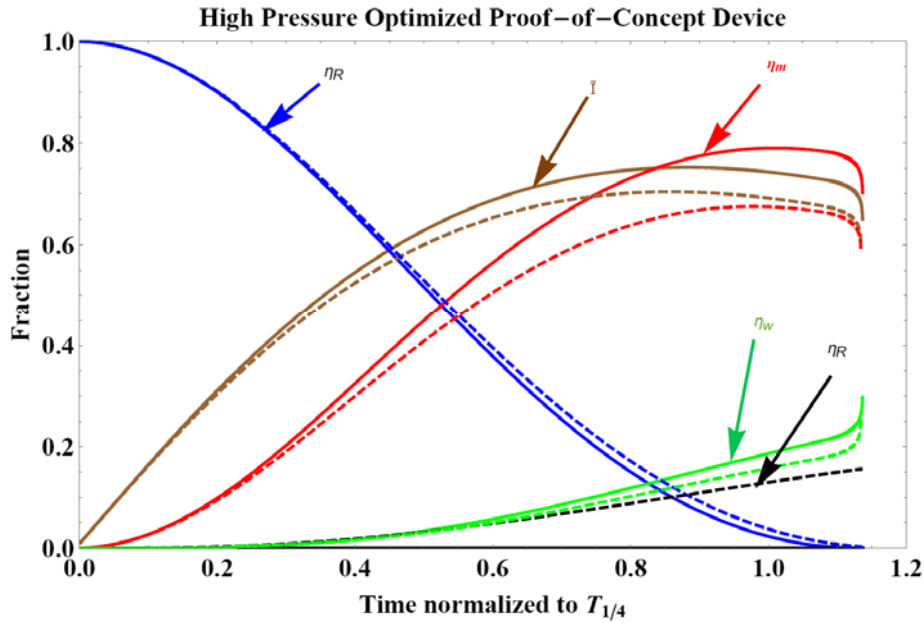

Fig 11: Energy conversion fractions and normalized current from RGV model[16] for the proposed High Pressure Optimized Proof-of-Concept system. Solid lines correspond to the $R_0=0$ case. Dashed lines are for $R_0=2$ mΩ case with reduced anode length shown in Fig 10..

Assuming a more realistic value of resistance is seen to delay the pinch as compared with the first order design iteration values obtained from the proposed algorithm. This delay is seen to be offset by reducing the anode length by trial. The resulting change in energy conversion scenario is seen to be tolerable in Fig. 11. Anode and cathode radii and anode length are seen to have practicable values. Insulator length, as mentioned above, is way too small: this requires innovative construction techniques.

The case of the High Pressure Optimized Industrial Scale device is quite similar. Fig. 12 and 13 show the corresponding results.

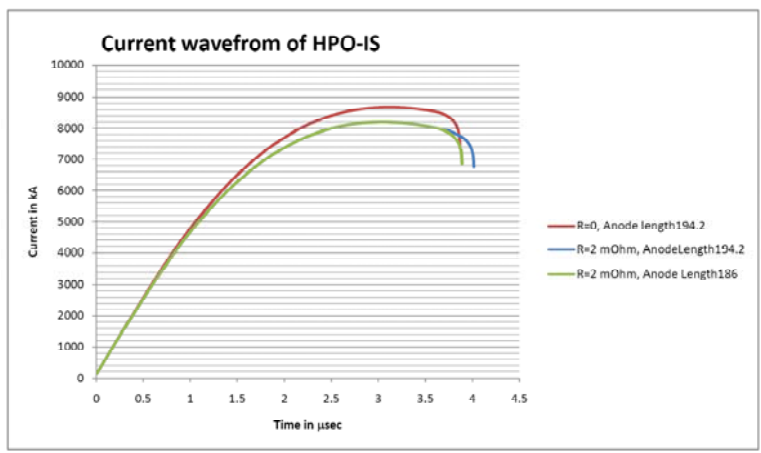

Fig 12: Current waveform calculated from RGV model [16] for proposed High Pressure Optimized Industrial Scale system. Initially, resistance is assumed to be zero. Assuming a more realistic resistance value of 2 mΩ leads to a delay in CDS. To compensate, anode length is reduced by trial.

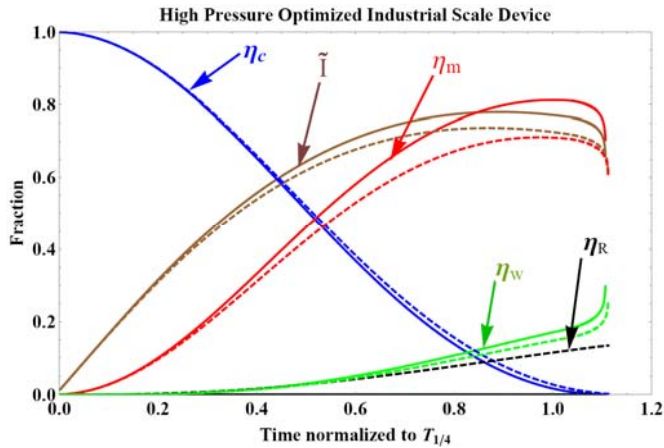

Fig 13: Energy conversion fractions and normalized current from RGV model[16] for the proposed High Pressure Optimized Industrial Scale system. Solid lines correspond to the $R_0=0$ case. Dashed lines are for $R_0=2$ m$\Omega$ case with reduced anode length as shown in Fig 12.

Figures 10-13 demonstrate that the proposed algorithm is able to converge to a set of first order high pressure optimized prototype design parameters that span a two orders of magnitude in energy and one order of magnitude in current and deuterium gas pressure. Significantly, both cases have drive parameters that fall within the Lee-Seban range [20].

The current waveform calculated from the RGV model can be treated as "measured current" which can be fitted to the Lee model [27]. Fig. 14 shows this fit for the PC system along with the fit parameter set. The Lee model reports neutron yield estimate as 7.81E+09 based on the phenomenological beam target mechanism.

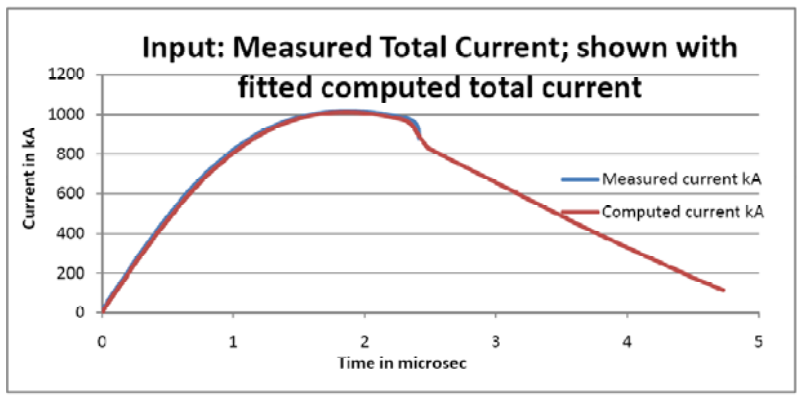

Fig 14: Fit of the current waveform calculated from RGV model for the proposed High Pressure Proof-of-Principle system with the Lee model. The Lee model parameter set is also displayed.

Fig 15 shows results of a similar exercise for the IS system. The Lee model estimates the neutron yield as 6.83E+13; the energy content of D-D fusion neutrons is ~27 Joules!

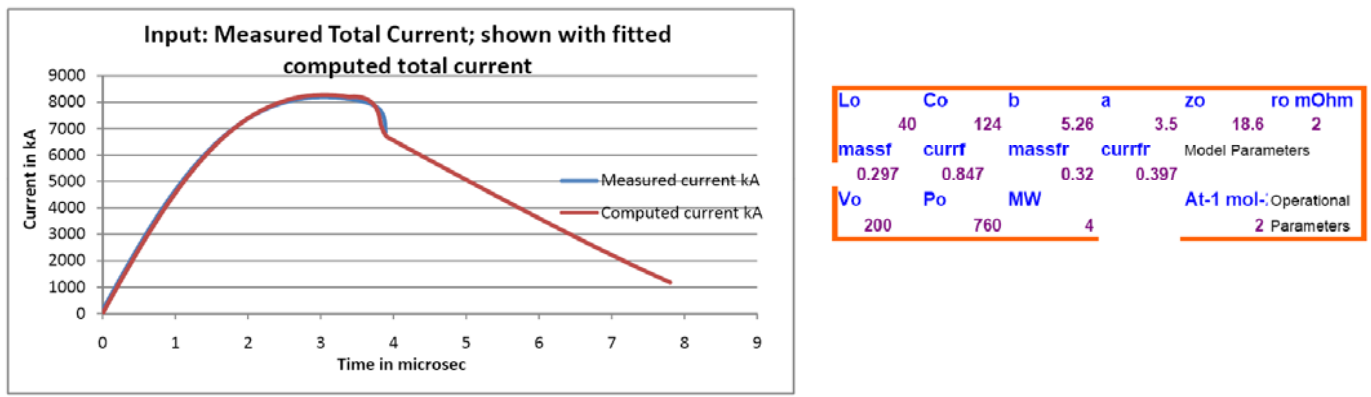

Fig 15: Fit of the current waveform calculated from RGV model for the proposed High Pressure Industrial Scale system with the Lee model. The Lee model parameter set is also displayed.

The Lee model fit of the current waveform computed using the RGV model shows that the two models are internally connected in spite of many differences in their structure and assumptions. Neutron yield estimates from the Lee model are empirically calibrated against data from existing devices. These estimates can be taken as confirmation that the proposed optimization algorithm does indeed lead to an optimized prototype system parameter set without any trial and error, that could reasonably (to the extent Lee model numerical experiments correspond with real experiments) be expected to result in good plasma focus action at a pressure one to two orders of magnitude above the conventional DPF operating pressure.

V. Discussion:

The Lee model estimates neutron yield from both thermonuclear and beam target processes and finds that the latter predominate. The neutron yield estimate from the beam target process calculated by Lee and Saw [18] is directly proportional to the square of the current flowing through the pinch at the start of the slow compression phase, to the square of the pinch length and to the number density (note that the density factor is missing in Ref 18 but has been corrected in a subsequent erratum), inversely proportional to the square-root of the maximum voltage and logarithmically dependent on the pinch radius, with a proportionality constant that is empirically calibrated against data from existing neutron optimized devices. An argument similar in spirit to the argument leading to the above result can be made in terms of the RGV model and the results of the hyperbolic conservation law approach to DPF[22].

Equation 47 of Ref [22] yields an estimate of the ion kinetic energy associated with radial, azimuthal and axial components of fluid velocity in the fully ionized plasma current sheath at the pinch:

$$\varepsilon(J) \sim \tfrac{1}{2} m_i v^2 \sim \frac{\mu_0}{4 n_0} \left( \frac{I_p}{2\pi r} \right)^2 \qquad 17$$

which is of order ~100 keV [22]. Since the binary nuclear reaction rate depends on the distribution of *relative velocities* of ion pairs, having a high average fluid velocity does not lead to reactions. When the plasma current sheath gets reflected from the axis, there is a reversal of radial component of velocity, but not of axial and azimuthal components, and mixing of the reflected plasma and incoming plasma in an interaction zone leads to an abrupt increase in the number of ion pairs having a high relative velocity ~v, which resembles an ion beam trapped in a target plasma. In this picture, there is no need for an instability-triggered plasma diode action to accelerate ions, in conformity with the fact that in many cases, there is no instability whatsoever and still the neutron yield is unaffected [28]. This is also in accordance with the neutron diagnostics[22] that reveals a toroidal motion of center of mass of reacting deuterons.

This kinetic energy is acquired by the plasma through the electromagnetic work done, which is estimated by the RGV model as $\eta_w E_0$. The total number of "beam" ions $N_b$ can therefore be estimated as

$$N_b \sim \frac{\eta_w E_0}{\frac{1}{2} m_i v^2} \qquad 18$$

Assuming the volume of the mixing zone to be V, the "beam flux" is

$$\Phi \sim v \frac{N_b}{V} \qquad 19$$

The number density of "target" ions is proportional to the fill gas number density $n_0$. The macroscopic reaction cross-section should therefore be

$$\Sigma \sim n_0 \sigma_{D-D}(\varepsilon/e)$$

So the reaction rate per unit volume should be

$$\Phi\Sigma \sim v \frac{N_b}{V} n_0 \sigma_{D-D}(\varepsilon/e)$$
$$\sim \frac{2\sqrt{2}\pi}{\sqrt{\mu_0 m_i}} \frac{\sigma_{D-D}(\varepsilon/e)}{V} \frac{\eta_w}{\tilde{I}_P} rL_0 I_0 n_0^{1.5} \qquad 20$$

Therefore the total reaction rate should be

$$\Phi\Sigma V \sim \frac{2\sqrt{2}\pi}{\sqrt{\mu_0 m_i}} \frac{\eta_w}{\tilde{I}_P} \{r\sigma_{D-D}(\varepsilon/e)\} L_0 I_0 n_0^{1.5} \qquad 21$$

In the spirit of the Lee model, one can interpret r as the radius of the piston. As r decreases, the relative kinetic energy $\varepsilon$ of ion pairs given by 17 increases but the number of ion pairs decreases, generating a pulse shape of limited duration related to the radial velocity resulting in a finite neutron yield.

An in-depth investigation of this argument is beyond the scope of this paper and reserved for another time. What is relevant to this paper is the expectation that the reaction rate should scale with the fill gas density as $\sim n_0^{1.5}$. This provides the motivation for the construction of high pressure optimized DPF devices.

## VI. Summary and conclusion:

This paper proposes an algorithm that is capable of generating a complete set of design parameters of an initial prototype of a dense plasma focus facility that efficiently transfers energy from the capacitor bank to the plasma at pinch time for fill gas pressures in the range of 0.1-1 bar of deuterium, in contrast with less than 10 mbar pressure commonly used in DPF research. The input to this algorithm consists of three discretionary variables: static inductance $L_0$, operating voltage $V_0$ and fill gas pressure p. Results of design parameter space survey over the range 10-50 nH, 10-200 kV and 0.1-1 bar of deuterium show some surprising features: higher pressures are seen to be more cost effective purely from energy transfer considerations without invoking any reaction mechanism and optimized insulator lengths are less than 4 mm in complete contrast with current research practice, indicating the need for research into new insulator configurations.

The use of this algorithm is illustrated using two numerical examples. One represents an attempt at proof-of-concept of high pressure operation at lowest possible cost, at stored energy 18 kJ and pinch current 0.875 MA, and the other represents an industrial scale facility operating at stored energy 2.5 MJ and pinch current 7.3 MA. The Resistive Gratton Vargas model code [16] is used to calculate the current waveform for the initial prototype parameter sets obtained from the algorithm and to iteratively arrive at more realistic parameters including circuit resistance and anode length. The parameter sets thus refined using the RGV model are used with the Lee model to estimate the neutron yield. High neutron yields estimated from Lee model are taken as vindication of the ability of the proposed algorithm to converge on to parameters of an initial prototype that optimize energy transfer efficiency at high operating fill gas pressure.

The Lee model uses an empirically calibrated phenomenological (predominantly) beam-target model for neutron yield estimation that assumes a plasma diode action triggered by instability as the source of ions. An argument similar in spirit to the derivation of the Lee-Saw[18] formula for beam-target yield is constructed using results from a hyperbolic conservation law approach to DPF that shows scaling of reaction rate as fill gas density to the power 3/2. Since only experiments can distinguish between the two phenomenological arguments, this provides strong motivation for pursuing research into high pressure optimized dense plasma focus.